\documentclass[12pt]{JHEP3} % 10pt is ignored! 
\usepackage{epsfig,multicol}
\usepackage{latexsym}
\usepackage{amssymb}
\usepackage{amsfonts}
\usepackage{amsmath}
\usepackage{bm}
\usepackage{times}

\title{Diffuse neutrino supernova background as a cosmological test}

\author{J. Barranco, Argelia Bernal, D. Delepine\\
Departamento de F\'isica, Divisi\'on de Ciencias e Ingenier\'ia, Campus Le\'on, 
Universidad de Guanajuato, Le\'on 37150, M\'exico
}

\date{\today} 
\abstract{The future detection and measurement of the diffuse neutrino supernova background will shed light on the rate of supernovae events in the Universe, the star formation rate and the neutrino spectrum from each supernova. Little has been said about what those measurements will tell us about the expansion history of the universe.  
The purpose of this article is to show that the detection of the diffuse supernova neutrino background will be a complementary tool for the study and possible discrimination of cosmological models.
In particular, we study three different cosmological models: the $\Lambda$ Cold Dark Matter model, the Logotropic universe and 
a bulk viscous matter-dominated universe. By fitting the free parameters of each model with the supernova Ia probe, we found that the predicted  number of events computed with the best fit parameters for the $\Lambda$-Cold dark matter model  and with the Logotropic model are the same, while  a bulk viscous matter-dominated cosmological model predicts $\sim 3$ times more events.  We show that the current limit set by Super-Kamiokande on the diffuse supernova neutrino background flux gives
complementary constraints on the free parameters of a bulk viscous matter-dominated universe. Furthermore, this limit implies, within a   $\Lambda$ Cold Dark Matter model, that the universe should be expanding with $H_0 > 21.5 ~\rm{Km/sec/Mpc}$ independently of the content of dark matter $\Omega_m$.}

\begin{document}
\section{Introduction}
It has been 30 years since the first observation of extragalactic neutrinos coming from the supernova (SN) 1987A  \cite{Hirata:1987hu,Bionta:1987qt}. The detection of a few neutrinos originated from this cataclysmic event ($\sim 19$ events) provided the source of uncountable works on neutrino physics and stellar dynamics showing the potential of this kind of events as a particle physics laboratories \cite{review}.
Therefore, invaluable information will be obtained through the neutrino detection of a future galactic supernova explosion. Nevertheless, the number of supernova explosions per galaxy is very small, perhaps one or two per century \cite{VanDenBergh:1991ke}.  In the meantime, while waiting for the next galactic SN, one may search for the cumulative neutrino flux from all past SN in the Universe 
\cite{Guseinov1967,Hartmann:1997qe,Ando:2004hc,Horiuchi:2008jz}. This flux has been called the Diffuse supernova neutrino background (DSNB) and it is a time independent, isotropic flux of neutrinos and anti-neutrinos that permeates all universe. 
Its detection would be easier through the inverse beta decay $\bar \nu_e+p\to n+e^{+}$, since the cross section of this process is two orders of magnitude bigger than all other neutrino interactions at the energy range relevant for DSNB \cite{Strumia:2003zx}. 
This flux has been searched in the past with negative results and SuperKamiokande has set an upper limit on diffuse electron antineutrinos 
$\phi_{\bar \nu_e}< 1.2~ \bar \nu_e s~ \textrm{cm}^{-2}\textrm{sec}^{-1}$ \cite{Malek:2002ns}. This limit  is close to the theoretical predictions and thus
DSNB detection will be inevitable as soon as future megaton neutrino detectors will be available \cite{Autiero:2007zj}. 
Once observed, DSNB will offer a complete picture of all supernova population of the universe and provide information about the 
supernova rate explosion, the neutrino spectrum from each supernova, neutrino properties and other exotica. 
In addition to those elements which are essential ingredients in the calculation of DSNB, the expansion rate of the universe is also needed which means that the flux depends on the cosmological model adopted \cite{Totani:1995rg,Ono:2007zza}. In the case of a canonical universe with Cold Dark Matter and a cosmological constant (a $\Lambda$-CDM universe), the DSNB depends strongly on $H_0$, the current Hubble constant, and only weakly on the matter density  of the universe  $\Omega_m$  and on the  cosmological constant density $\Omega_\Lambda$ \cite{Totani:1995dw}. 
The purpose of this article is to study to what extent the DSNB flux changes depending of the cosmological model used. This work is not exhaustive and we consider only two alternative cosmological models, namely: a Logotropic universe \cite{Chavanis:2015paa,Chavanis:2015eka,Chavanis:2016pcp} and a universe without cosmological constant but with dark matter with volumetric bulk viscosity \cite{Avelino:2008ph,Avelino:2010pb}. We will show that for the latter case, the DSNB can change significantively with respect to the predicted flux by a $\Lambda$-CDM model or a Logotropic universe. 
Furthermore we will show that the present limit measured by Super-Kamiokande can give stronger constraints on the free parameters for the model with volumetric bulk viscosity. 

The main contribution to the total DSNB comes from SN at low redshift ($z<1$), and thus, DSNB is mainly sensitive to the expansion of the local universe. Although it seems as a throwback, this fact suggest that DSNB could be a cosmological tool for local measurements. 
Given the current tension between the direct local measurement of the Hubble parameter $H_0$ and the model dependent value inferred  from the cosmic microwave background \cite{Bernal:2016gxb,Riess:2016jrr}, the future measurement of DSNB could shed light on this controversy.

The article is organized as follows: In section \ref{sec1}, we describe briefly the main ingredients needed to calculate the DSNB. Then we compute the DSNB flux  and the number of events for a typical 22 kiloton detector within the $\Lambda$-CDM. 
In section \ref{sec2} we explore two alternative models: the Logotropic universe \cite{Chavanis:2015paa,Chavanis:2015eka,Chavanis:2016pcp} and a matter dominated universe with bulk viscosity \cite{Avelino:2008ph,Avelino:2010pb}.
Each model introduces different free parameters and since we are interested in  the local measurement of the expansion of the universe, we  fit those free parameters by means of a $\chi^2$ function using the Union 2.1 data set of the distance moduli of the observed Supernova \cite{Suzuki:2011hu}.
Once we have fixed those free parameters, we compute the DSNB and the number of events in section \ref{sec3}.
We found that the Logotropic 
universe predicts the same number of events as $\Lambda$CDM, while the  bulk viscous matter-dominated universe exceed this number by a factor $\sim 3$. Then by using the Super-Kamiokande limit on DSNB we set bounds on the free parameters of the Bulk viscous model. 
Finally, some conclusion and perspectives are commented in section \ref{sec4}.

\section{Diffuse neutrino supernova background}\label{sec1}
A core collapse supernova (CCSN) explosion is one of the most energetic events in astrophysics. About $99\%$ of its gravitational binding energy is transformed into neutrinos (around $10^{53} $ ergs).  
Along the history of the universe, many CCSN explosions have been occurred and the cumulative emission of from all past supernova forms a diffuse neutrino or antineutrino background. This is the so called Diffuse supernova neutrino background (DSNB). In this section we will briefly explain how to compute DSNB and its event spectrum, and show how DSNB depend on the cosmological model adopted. 
 
\subsection{Computing the DSNB}

The neutrino or antineutrino component of DSNB  can be computed by integrating the core collapse supernova rate $R_{CCSN}$
multiplied by the neutrino or antineutrino emission spectrum $\frac{dN(E)}{dE}$ over the cosmic time, i.e.  \cite{Ando:2004hc,Raffelt:2009mm}
\begin{equation}
\frac{d\phi^{DSNB}}{dE}=\int R_{CCSN}(z)\frac{dN(E)}{dE}\big{|}\frac{dt}{dz}\big{|}dz, \label{flujo}
\end{equation}

where 
\begin{itemize}
\item $R_{CCSN}$ is the core collapse SN rate. This is the rate of supernova explosions per unit of comoving volume. This rate is 
proportional to the star formation rate (SFR) $\psi_*(z)$ times the fraction of stars with masses bigger that $8M_\odot$ and assuming that 
all such stars undergo CCSN explosion.  SFR is derived from measurement of the luminosities of massive stars which in addition to the 
knowledge of their masses and their life times gives their birth rates.  Following \cite{Horiuchi:2008jz,Yuksel:2008cu}, the SFR function can be
parametrize by a broken power-law given by:
 
\begin{equation}
\psi_*(z)=\dot \rho_0\left[(1+z)^{\alpha \eta}+\left(\frac{1+z}{B}\right)^{\beta \eta}
+\left(\frac{1+z}{C}\right)^{\gamma \eta}\right]^{1/\eta}, \label{rccsn}
\end{equation} 
where the constants $\alpha$, $\beta$, $\gamma$ and $\eta$ together with $\dot \rho_0$ are obtained through a parametric fit of the redshift evolution of the comoving  star formation rate density. Using the data compilation of Hopkins and Beacon \cite{Hopkins:2006bw}  it was found  that $\dot \rho_0=0.0178^{+0.0035}_{-0.0036} M_{\odot}$ yr$^{-1}$ Mpc$^{-3}$, $\alpha=3.4 \pm 0.2$, $\beta=-0.3 \pm 0.2$, 
$\gamma=-3.5 \pm 1.$. Uncertainties refer to an upper and lower envelope that takes into account the scatter in the data \cite{Horiuchi:2008jz}.
Furthermore,  $B=(1+z_1)^{1-\alpha/\beta}$ and $C=(1+z_1)^{(\beta-\alpha)/\gamma} (1+z_2)^{1-\beta/\gamma}$ with $z_1=1$ and $z_2=4$ \cite{Horiuchi:2008jz}.
Given this SFR, the core collapse SN rate can be computed as 
\begin{equation}
R_{CCSN}(z)=\frac{\int_8^{50} \psi(M) dM}{\int_{0.1}^{100}M \psi(M) dM} \psi_*(z)=\frac{0.007}{M_{\odot}}\psi_*(z)
\end{equation}
where $\psi(M)$ is the initial mass function (IMF) and it was assumed a Salpeter function. 
Different parametrizations for $\psi_*(z)$ has been done, but the fit is obtained from astrophysical measurements and thus, it is independent of the 
assumed cosmological model.  

\item $\frac{dN(E)}{dE}$ is the time integrated neutrino spectrum per SN. In the Kelvin-Helmholtz cooling fase of a core collapse SN neutrinos and antineutrinos are produced. Numerical simulations of core collapse SN explosions predicts a neutrino or antineutrino spectrum which can be fitted by
\cite{Keil:2002in}:
\begin{equation}
\frac{dN(E)}{dE}=\frac{(1+\alpha)^{1+\alpha}E_{\rm tot}}
{\Gamma(1+\alpha)\bar E^2}
\left(\frac{E}{\bar E}\right)^{\alpha}e^{-(1+\alpha)E/\bar E}\,.
\end{equation}
Here  $\bar E$ is the average energy of the neutrino or antineutrino. For the antineutrino case, since this is the flux we are interested in computing because the large cross section for a possible detection of DSNB, the fitting constants are $\alpha=4$, $E_{\rm tot}=5\times10^{52}$~erg, which is the total energy emitted \cite{Keil:2002in}. 

\item Finally $\big{|}\frac{dt}{dz}\big{|}$ is related to the Hubble parameter $H(z)$ trough $|\frac{dz}{dt}|=(1+z)H(z)$. $H$ is given by 
the Friedman equation: 
\begin{equation}
H^2=\left(\frac{\dot a}{a}\right)^2=\frac{8\pi G}{3 c^2}\rho\,, \label{hubble}
\end{equation} 
where $a$ is the scale factor and $\rho$ is the energy density. Here we have assumed zero curvature. 

In a standard cosmological model, the expansion of the universe is induced by a cosmological constant $\Lambda$ and there is a pressureless
matter that interacts only by its gravitational force with the baryonic matter. 
The energy density has two contributions: 
One given by this non interacting matter. i.e. dark matter, and the energy density given by the cosmological constant. Thus 
$\rho=\rho_m+\rho_\Lambda$ and
\begin{equation}
\dot \rho_m+3 H\left( \rho_m+\frac{p}{c^2}\right)=0\,, \label{fluid}
\end{equation}
gives the evolution of the mass density. In this case the matter is pressureless and thus $p=0$. This is the so-called $\Lambda$-CDM model. 
Solving eqs. \ref{hubble}-\ref{fluid}, the Hubble parameter for $\Lambda$-CDM model is:
\begin{equation}
H(z)=H_0 \sqrt{\left(\Omega_\Lambda+(1+z)^3 \Omega_m\right)}\,,\label{HLCDM}
\end{equation}
where $\Omega_\Lambda=\rho_\Lambda/\rho_c$, $\Omega_m=\rho_m/\rho_c$ and $\rho_c=\frac{3H^2}{8\pi G} \sim 10^{-26}\rm{Kg}/\rm{m}^3$. 
The values of $\Omega_m,\Omega_\Lambda$ and $H_0$ have been fixed by astrophysical observations, namely: the observational evidence from supernova for an accelerating universe \cite{Riess:1998cb,Perlmutter:1998np} and the cosmic background anisotropies \cite{Ade:2015xua}.  

\end{itemize}

%\FIGURE{\epsfig{file=flujoDSNB.eps,width=5cm} 
%        \epsfig{file=eventos_cosmology.eps,width=5cm} 
%        \caption[Left panel: Predicted diffuse supernova neutrino background flux for a $\Lambda$-CDM model with the best fit parameters as measured by Planck collaboration, 
%i.e. $H_0=68.2$~Km/sec/Mpc, $\Omega_m=0.28,\Omega_\Lambda=0.72$. Upper and lower lines for the flux that gives the grey area in the DSNB spectrum are obtained by using the upper and lower 
%envelopes that takes into account the scatter in the data of the
%core collapse supernova rate $R_{CCSN}$ \cite{Horiuchi:2008jz}. Right panel: Predicted number of events for the DSNB flux form the left for a 22.5 KTon water detector according to eq. \ref{events} with the same parameters for the flux as shown in the left panel.  The dashed red line is the number of events for the value of the Hubble constant as measured by local data, i.e. $H_0=73.24$ Km/sec/Mpc.]}%
%	\label{flujoLCDM}}

\begin{center}
%\begin{figure}
\FIGURE{
\includegraphics[width=0.45\textwidth]{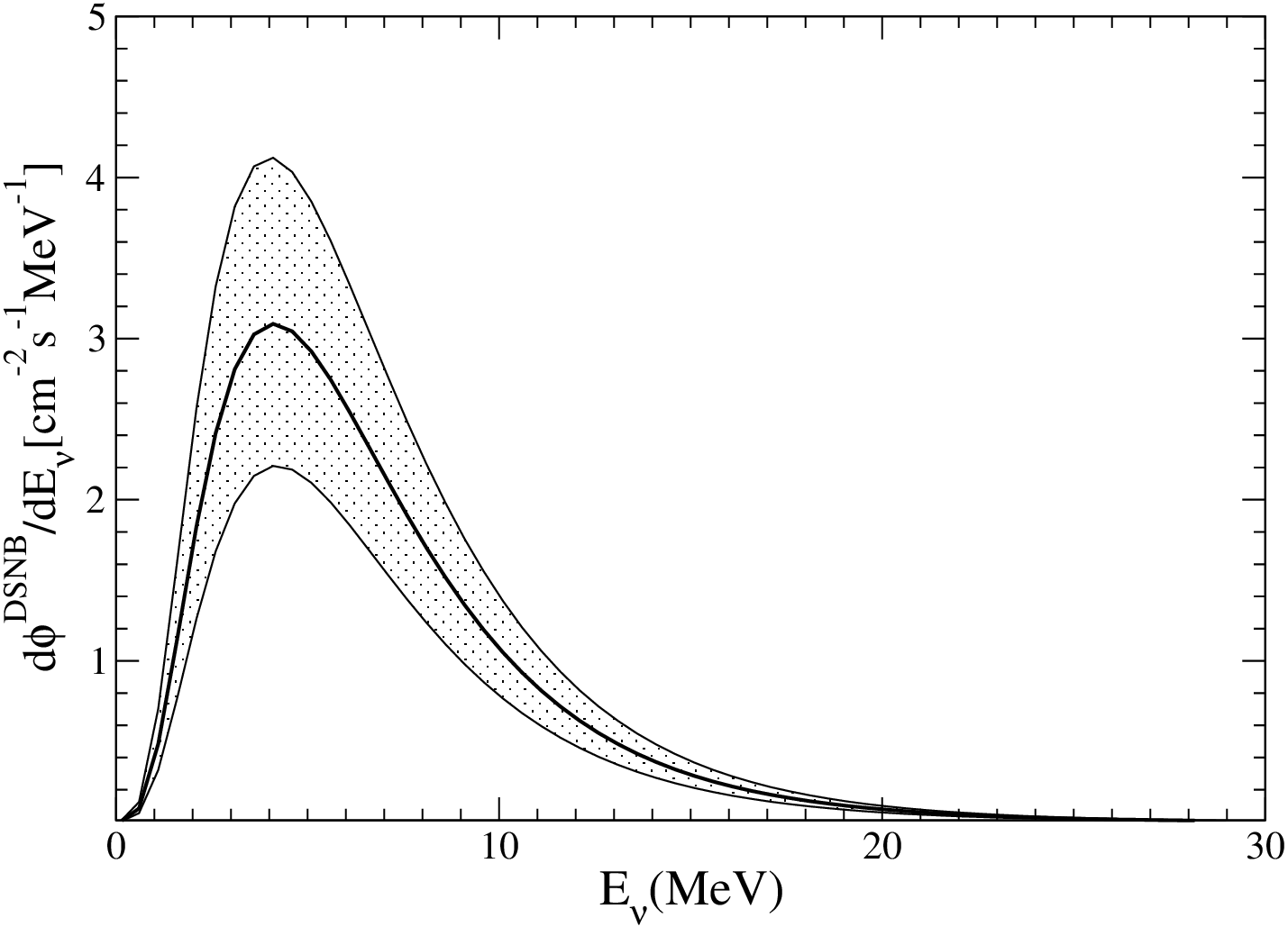}
\includegraphics[width=0.45\textwidth]{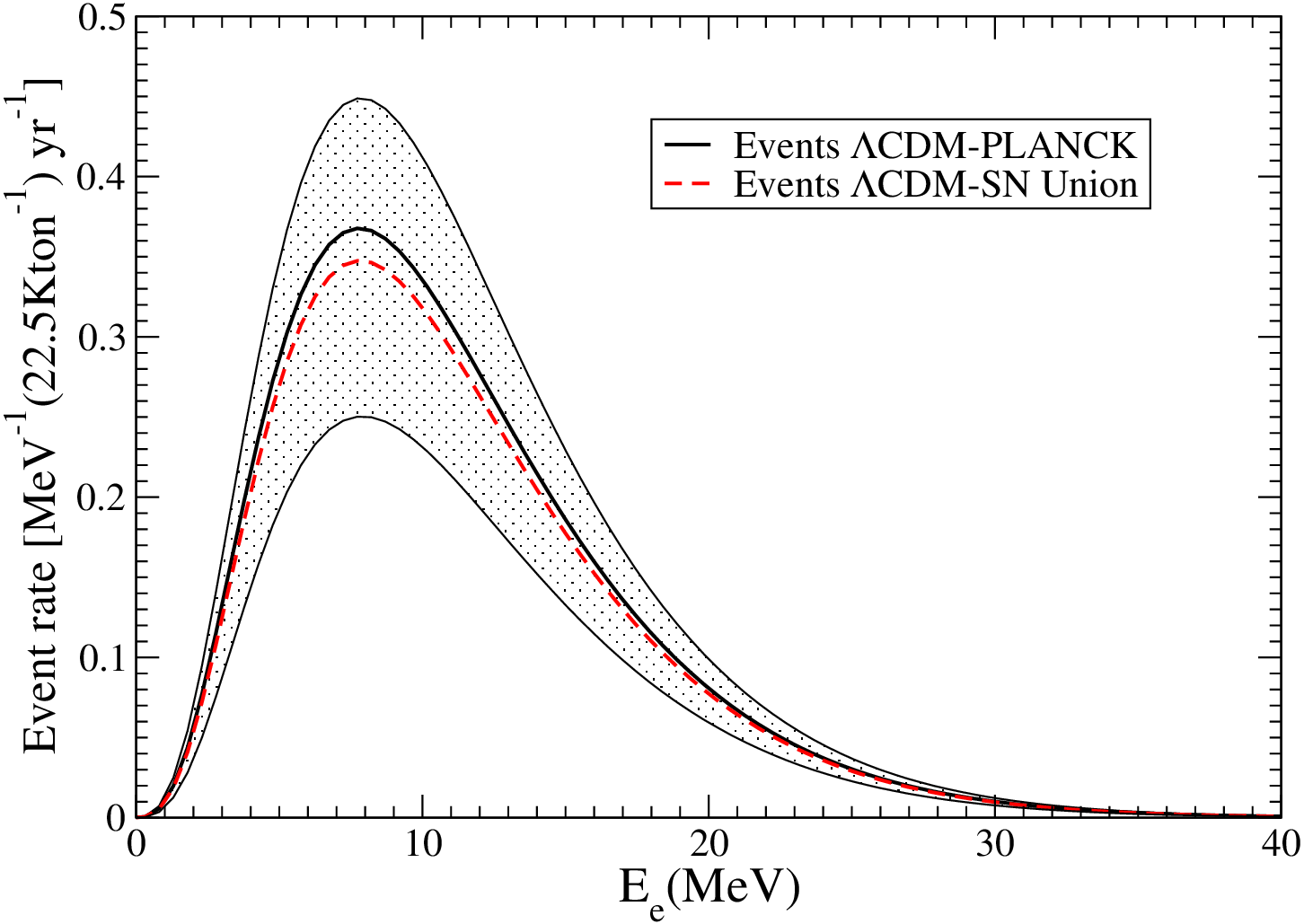}
\caption{Left panel: Predicted diffuse supernova neutrino background flux for a $\Lambda$-CDM model with the best fit parameters as measured by Planck collaboration, 
i.e. $H_0=68.2$ Km/sec/Mpc, $\Omega_m=0.28,\Omega_\Lambda=0.72$. Upper and lower lines for the flux that gives the grey area in the DSNB spectrum are obtained by using the upper and lower 
envelopes that takes into account the scatter in the data of the
core collapse supernova rate $R_{CCSN}$ \cite{Horiuchi:2008jz}. Right panel: Predicted number of events for the DSNB flux form the left for a 22.5 KTon water detector according to eq. \ref{events} with the same parameters for the flux as shown in the left panel.  The dashed red line is the number of events for the value of the Hubble constant as measured by local data, i.e. $H_0=73.24$ Km/sec/Mpc.}\label{flujoLCDM}
%\end{figure}  
}
\end{center}

Taking into account those three key ingredients, integration of Eq. \ref{flujo} gives the DSNB flux. The result is shown in left panel of Fig. \ref{flujoLCDM}.
The biggest uncertainty comes from the rate of supernova explosions per unit of comoving volume $R_{CCSN}$. We have included this uncertainty by varying the parameters 
$\dot \rho,\alpha, \beta$ and $\gamma$ in eq. \ref{rccsn} such as we include  an upper and lower envelope that takes into account the scatter in the data \cite{Horiuchi:2008jz}.

Now we can predicted the event rate spectrum. It is estimated as the flux spectrum weighted with the detection cross section $\sigma (E_\nu)$ 
\begin{equation}
\frac{dN}{dE_e}=N_p \sigma (E_\nu)   \frac{d\phi^{DSNB}}{dE_\nu},\label{events}
\end{equation}
where $N_p$ is the number of proton targets, $E_e$ the energy of the positron. 
The cross section of the inverse beta decay $\bar \nu_e+p\to n+e^{+}$  is two orders of magnitude bigger than other
neutrino interactions in the energy regime of interest for DSNB flux. We  use the cross section  as reported \cite{Strumia:2003zx}.
\begin{center}
\FIGURE{
%\begin{figure}
\includegraphics[width=0.9\textwidth]{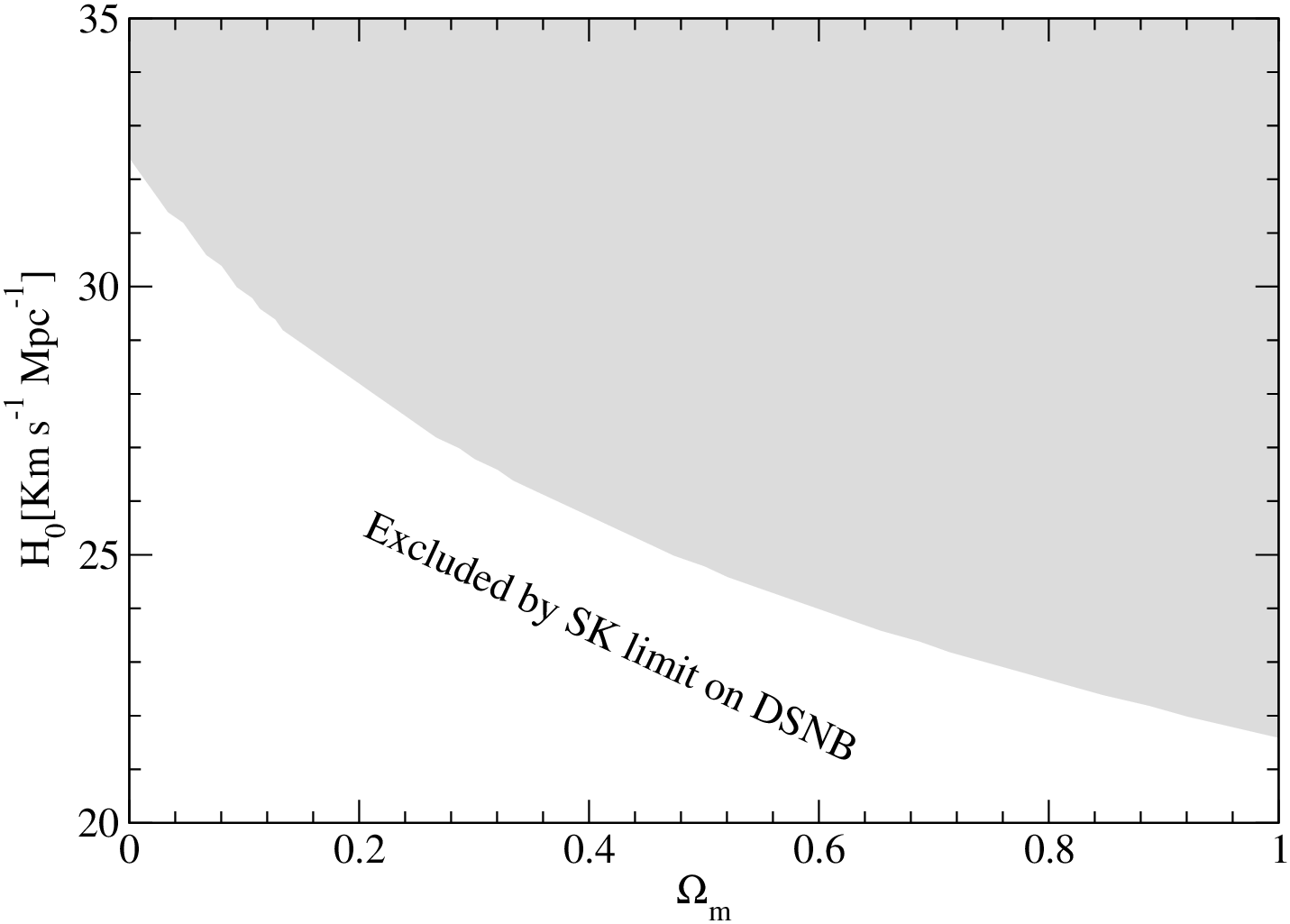}
\caption{Allowed region for  $H_0,\Omega_m$ to fulfill SuperKamiokande limit on DSNB that implies  $H_0 > 21.5 ~\rm{Km/sec/Mpc}$ independently of the content of dark matter $\Omega_m$. }\label{LimitCDM}
}
%\end{figure}  
\end{center}
In the right panel of  Fig. \ref{flujoLCDM} we show the expected number of events for a  $\Lambda$CDM scale factor $H(z)$ (see eq. \ref{HLCDM}). 
The uncertainty reported as the grey area is obtained as explained for the DSNB flux. Solid black line is the predicted number of events for $H_0=68.2 ~\rm{Km/sec/Mpc},\Omega_m=0.28,\Omega_\Lambda=0.72$ as reported by the Planck collaboration. Given the current controversy on the determination of $H_0$ \cite{Riess:2016jrr}, we have included the red dashed line for the predicted number of events  with the value of $H_0=70. ~\rm{Hm/sec/Mpc}$ as obtained by fitting $H_0$ using the SN Ia data from the "Union 2.1" data set.
As expected, given the higher value on $H_0$  the number of events decreases. Unfortunately, astrophysical uncertainty is bigger that the possible miss-match in the predicted number of events obtained with "local" determination of the cosmological parameters and the ones obtained by the cosmic microwave anisotropies. 
\subsection{DSNB current limit and implications for $\Lambda$-CDM}
The results shown in Fig. \ref{flujoLCDM} indicate that the measurement of the DSNB will be inevitable in the near future with future megaton neutrino detectors \cite{Autiero:2007zj}. Currently an upper limit on the DSNB was obtained by looking for electron-type antineutrinos that had produced a positron using 1496 days of data from the Super-Kamiokande detector. The non observation of events implies an upper limit of $1.2~ \bar \nu_e \rm{cm}^{-2} \rm{sec}^{-1}$ for antineutrinos with energy $E_\nu > 19.3$ MeV \cite{Malek:2002ns}. 
Interestingly, this limit already can gives us some insight about the rate of expansion of the universe. Indeed, we can compute the DSNB flux as explained above using eq. \ref{flujo} for a $\Lambda$-CDM model ($\Omega_\Lambda=1-\Omega_m$). We can use the upper values of $\dot \rho_0,\alpha,\beta,\gamma$ which give the highest DSNB flux for a given set of values for $H_0$ and $\Omega_m$. Thus, the DSNB flux will be a function of  $H_0$ and $\Omega_m$, i.e. $\frac{d\phi^{DSNB}}{dE}$ ($H_0,\Omega_m$), and we can integrate the total flux for $E_\nu> 19.3$ MeV. Thus, by demanding $\phi^{DSNB}(H_0,\Omega_m) < 1.2~ \bar \nu_e \rm{cm}^{-2} \rm{sec}^{-1}$ we obtain the region shown in Fig. \ref{LimitCDM}. 
As it can be seen, the current limit of SuperK implies that the universe should be expanding with $H_0 > 21.5 ~\rm{Km/sec/Mpc}$ independently of the content of dark matter $\Omega_m$.
\begin{center}
\FIGURE{
%\begin{figure}
\includegraphics[width=0.45\textwidth]{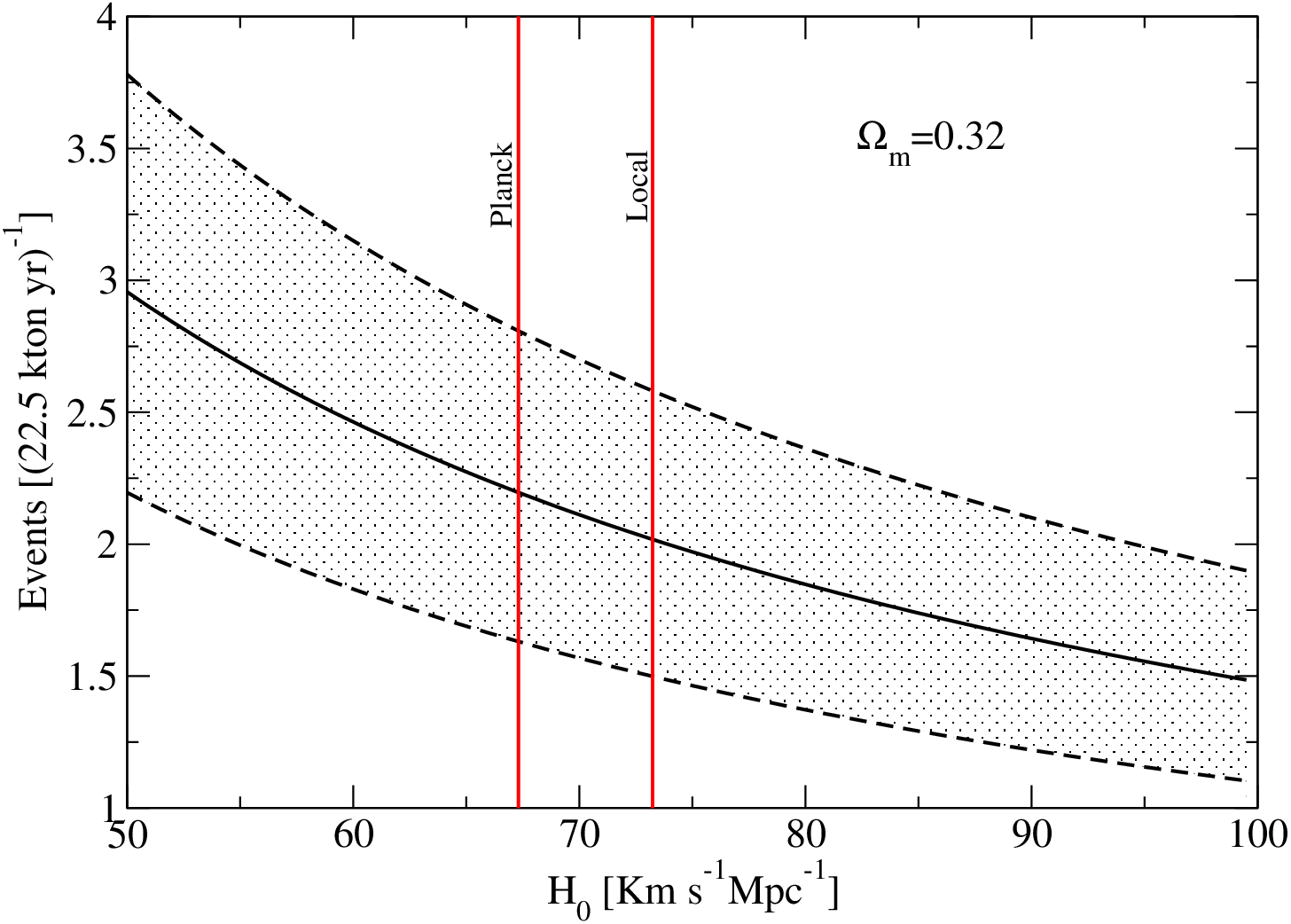}
\includegraphics[width=0.45\textwidth]{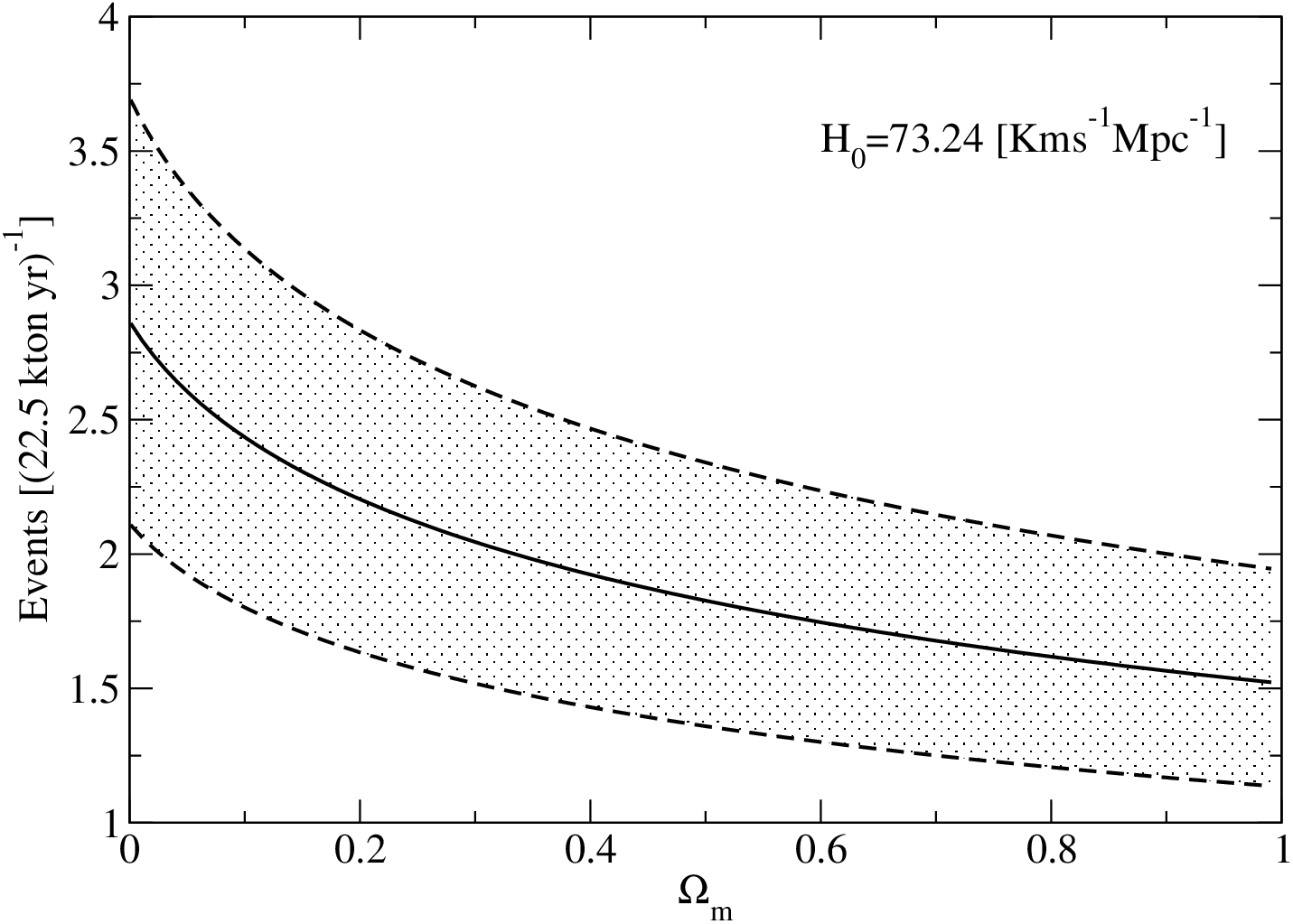}
\caption{The events for a 22 kiloTon detector with Gadolidium}\label{perspectives}
}
%\end{figure}  
\end{center}
A promising technique proposed in \cite{Beacom:2003nk} might improve significantly the DSNB detection. The idea is to dissolve GdCl$_3$ into the water of Super-Kamiokande. It is well known that the neutron capture efficiency of Gd is estimated to be $90\%$ in a $0.2\%$ admixture of GdCl$_3$ which will enhance the energy window of detection of Super Kamiokande to the range $E_e \in $[10-30] MeV.  Integration of eq. \ref{flujo} in this energy range gives the total number of events as a function of $H_0$ (left panel Fig. \ref{perspectives}) or as a function of $\Omega_m$ (right panel Fig. \ref{perspectives}).  The grey area between the two dashed lines corresponds again to the uncertainty in the $R_{CCSN}$. 
In order to disentangle the controversy on $H_0$, future experiments must reduce its error below a $5\%$ which represents the difference between the predicted number of events for a 22.5 Kton detector enriched with $\rm{GdCl}_3$ for $H_0=73.25$Km/sec/Mpc as stablished with local astrophysical measurements of the accelerated expansion of the universe and  $H_0=68.$Km/sec/Mpc as inferred from Planck data.

\section{Non standard Cosmological Models}\label{sec2}

%\begin{figure}
\begin{center}
\FIGURE{
\includegraphics[angle=0,width=0.75\textwidth]{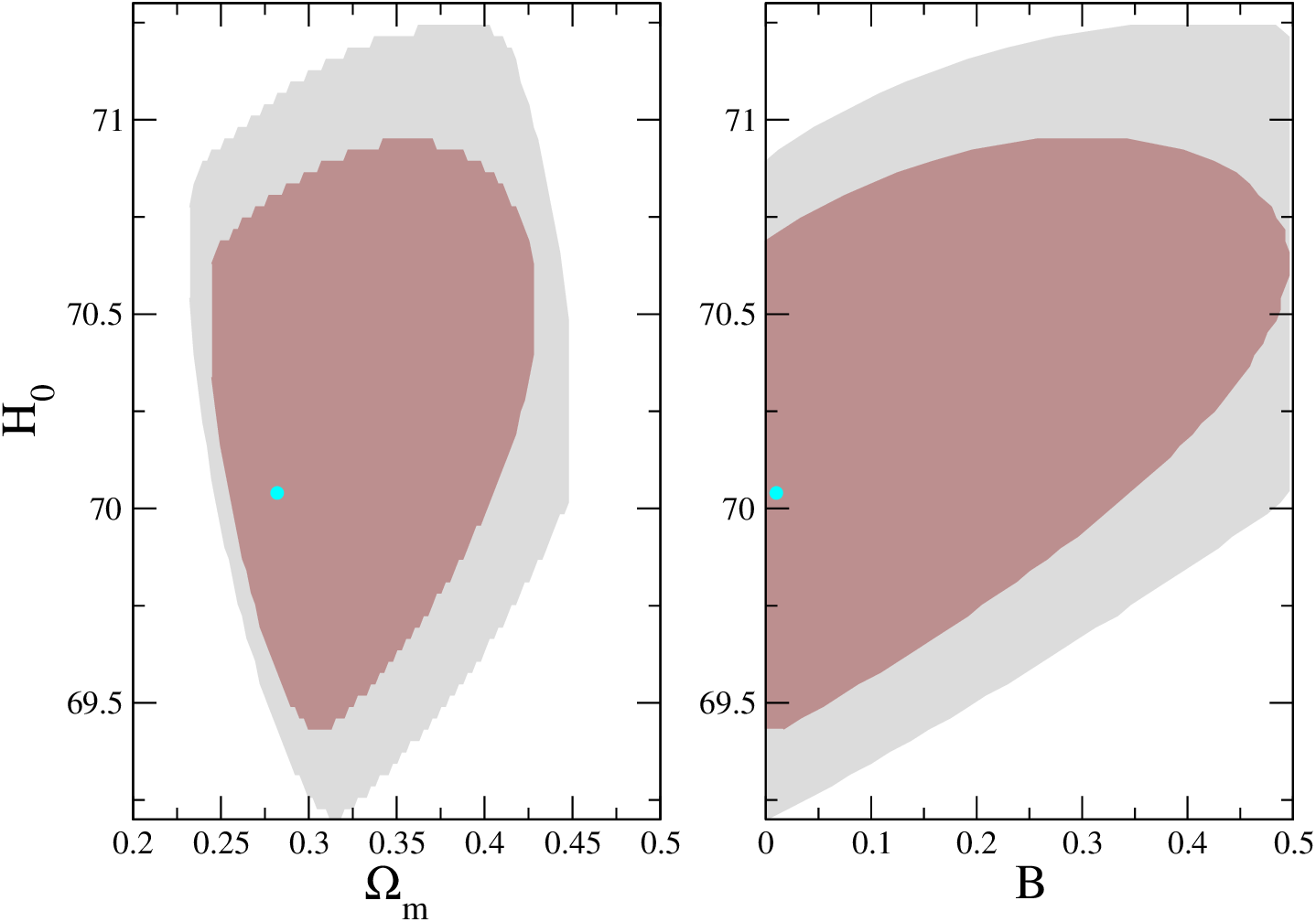}
\caption{Contraints on the free parameters of the Logotropic Cosmological model. Cyan points represent the best fit point and the colored areas are the allowed region for the free parameters at $68\%$ C.L (brown) and $90\%$ C.L. (grey) obtained by
minimizing the $\chi^2$ function (eq. \ref{chi2}) over the 580 points of the Union 2.1 data set}\label{cosmology_fit1}
}
\end{center}
%\end{figure}  

The cosmological constant, which correspond to the energy density of the vacuum, is the simplest model to explain an accelerating expansion of the universe.  Nevertheless, two problems arise:
\begin{enumerate}
\item a huge discrepancy between its predicted value and the observed value  and 
\item the ``\emph{cosmic coincidence problem}'', i.e. the problem that we are living in a time when the
matter density in the Universe is of the same order than the dark energy density.
\end{enumerate} 
In order to solve these problems, several models have been introduced. In what follows, we will mention two alternative cosmological models. Then,
we test its viability by performing a $\chi^2$ statistical analysis using the "Union" data set and the free parameters are constrained. 
We compute the predicted event rate of supernova relic neutrinos for each cosmological model. Finally, by using the current upper limit 
on the antineutrino diffuse supernova neutrino background set by SuperKamiokande, we show that it is possible to constrain
some alternative models showing the complementary of the DSNB and other cosmological observations, specially at low redshifts.
\begin{center}
\FIGURE{
\includegraphics[angle=0,width=0.75\textwidth]{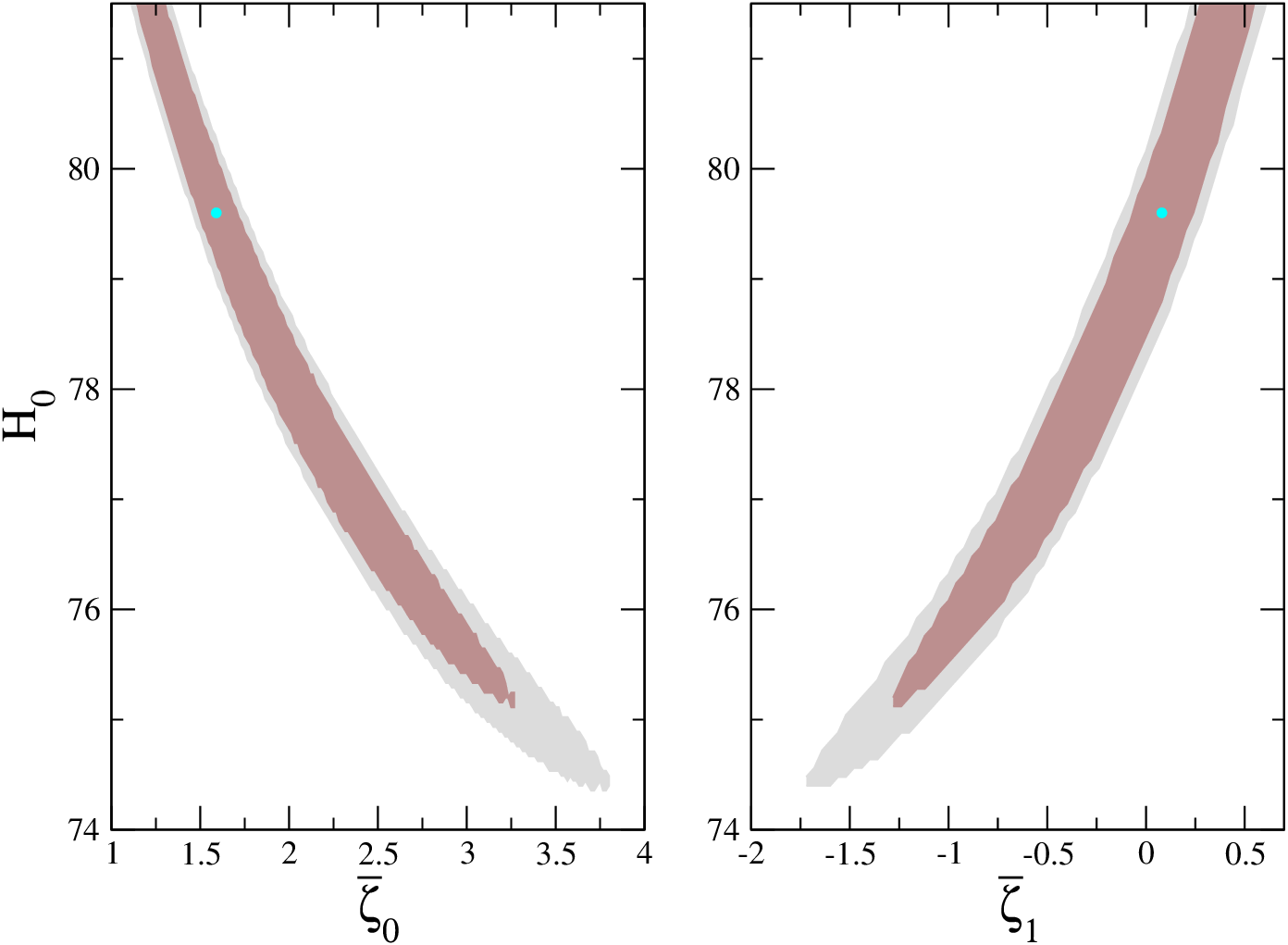}
\caption{Contraints on the free parameters of the volumetric viscous cosmological model.
Cyan points represent the best fit point for each model, and the colored areas are the allowed region for the free parameters at $68\%$ C.L (brown) and $90\%$ C.L. (grey) obtained by
minimizing the $\chi^2$ function (eq. \ref{chi2}) over the 580 points of the Union 2.1 data set}\label{cosmology_fit2}
}
\end{center}

\subsection{Logotropic Universe}
In this model dark matter and dark energy are unified in a single fluid with a logotropic equation of state 
\begin{equation}
P=A \log (\rho/\rho_P)\,, \label{EOS_logo}
\end{equation}
where $\rho$ is the rest mass energy, $\rho_P$ is the reference density, and $A$, which is  a free parameter of the model.  
Integration of the first law of thermodynamics for an adiabatic evolution of a perfect fluid with an equation of state of the form given by eq. \ref{EOS_logo}
gives for the energy density $\epsilon$ \cite{Chavanis:2015paa}
\begin{equation}\label{energy_density}
\epsilon = \rho c^2 -A\ln\left(\frac{\rho}{\rho_P}\right)-A\,,
\end{equation}
that may be written as
\begin{equation}
\epsilon = \rho_m c^2 + \rho_{DE}\,,
\end{equation}
\begin{equation}
\rho_{DE}=-A\ln\left(\frac{\rho}{\rho_P}\right)-A\,. \label{rhode}
\end{equation}
The evolution of $\rho_m$ evolves as $\frac{\rho_{m}}{\rho_c}=\Omega_{m0} a^{-3}$ and  the dark energy density 
$\frac{\rho_{DE}}{\rho_c}=\Omega_{DE0}+3\frac{A}{\rho_c} \ln a$ where $a$ is the scale factor and where the subscript `$0$' again refers to the present time.
Inserting this in eq. \ref{hubble}, thus, the Hubble parameter considering this logotropic fluid in terms of the redshif $z$ is given by
\begin{equation}
H(z)=H_0\sqrt{\Omega_{m0}(1+z)^3+(1-\Omega_{m0})(1-3 B \log(1+z))}\,.\label{hubble_logo}
\end{equation}
where $B=A/\rho_{DE}$ which is known as the logotropic temperature. 

\subsection{Bulk viscous matter-dominated universe}
A cosmological model with a pressureless fluid with a constant bulk viscosity can be an explanation of the accelerated expansion of the universe. For this fluid, 
the energy-momentum tensor of an imperfect fluid can be expressed as \cite{Avelino:2008ph,Avelino:2010pb}:
\begin{equation}\label{tensor_bulk}
T_{\mu\nu}=\rho_{\rm m} \,u_\mu u_\nu + (g_{\mu\nu}+u_\mu u_\nu)P^*_{\rm m}\;,
\end{equation}
here $u^\mu$ is the four-velocity vector of an  observer who measures the energy density $\rho_{\rm m}$, $g_{\mu\nu}$ is the metric tensor and  
there is an effective pressure $P^*_{\rm m}$ which is given in terms of the pressure of the fluid of matter $P_{\rm m}$  and it is affected by the bulk viscosity $\zeta$. 
Here we follows \cite{Avelino:2008ph,Avelino:2010pb} and we write this effective pressure as:
\begin{equation}
P^*_{\rm m} \equiv P_{\rm m} - \zeta \nabla_{\nu}u^{\nu}\,.
\end{equation}
Conservation equation for this viscous fluid gives
\begin{eqnarray}\label{equation1PerfectFluid}
u^{\nu} \nabla_{\nu} \rho_{\rm m} + (\rho_{\rm m} + P^*_{\rm m}) \nabla_{\nu} u^{\nu} &=0\,.
\end{eqnarray}
Assuming a spatially flat geometry for the Friedmann-Robertson-Walker (FRW) cosmology, this conservation equation can be rewritten as
\begin{eqnarray}
\dot{\rho}_{\rm m}  + 3H(\rho_{\rm m}  + P^*_{\rm m}) &=0\,,  \label{conservation}
\end{eqnarray}
where $H \equiv \dot{a}/a$ is the Hubble parameter as usual.

From a phenomenological point of view, let us assume the following ansatz for the viscosity coefficient $\zeta=\zeta_0+\zeta_1 H$.
Thus, rewriting the conservation equation in terms of the scale factor we get
\begin{equation}
\frac{d \rho_{\rm m}}{da} + \frac{(3-\bar{\zeta}_1)}{a}\rho_{\rm m} - \frac{3H_0}{(24 \pi G)^{1/2}} \frac{\bar{\zeta}_0}{a}\rho_{\rm m}^{1/2} = 0\,.
\end{equation}
Normalizing by the critical density today
$\rho^0_{\rm crit} \equiv 3H^2_0 / 8\pi G$ and defining  the dimensionless bulk viscous coefficients $\bar{\zeta}_0$ and $\bar{\zeta}_1$ as
\begin{equation}
\bar{\zeta}_0 \equiv \left(\frac{24\pi G}{H_0}\right) \zeta_0, \qquad \bar{\zeta}_1 \equiv \left(24\pi G\right) \zeta_1\,, \label{zetas}
\end{equation}
it is finally obtained
\begin{equation}\label{hubbleviscosoa}
a\frac{d \hat{\Omega}_{\rm m}}{da} + (3 - \bar{\zeta}_1) \hat{\Omega}_{\rm m} -
\bar{\zeta}_0 \hat{\Omega}_{\rm m}^{1/2} =0\,,
\end{equation}
where as usual ${\Omega}_{\rm m}\equiv \rho_{\rm m}/\rho^0_{\rm crit}$. 
In terms of the redshift $1+z= 1/a$, eq.  \ref{hubbleviscosoa}  can be rewritten as

\begin{equation}\label{hubbleviscosoz}
(1+z)\frac{d {\Omega}_{\rm m}}{dz} +(\bar{\zeta}_1-3){\Omega}_{\rm m} + \bar{\zeta}_0{\Omega}_{\rm m}^{1/2}=0\,.
\end{equation}
This equation has as solution
\begin{equation}\label{solution}
{\Omega}_{\rm m}(z) = \left[ \left(1 -\frac{\bar{\zeta}_0}{3-\bar{\zeta}_1} \right) (1+z)^{(3-\bar{\zeta}_1)/2} +\frac{\bar{\zeta}_0}{3-\bar{\zeta}_1} \right]^2\,.
\end{equation}
Here  $\Omega_{\rm m 0}={\Omega}_{\rm m}(z=0) =1$.
Now it is straighforward to write $H(z)$ as
\begin{equation}
H(z)=H_0\left(\left(1-\frac{\bar \zeta_0}{3-\bar \zeta_1}\right)(1+z)^{(3-\zeta_1)/2}+\frac{\bar \zeta_0}{3-\bar \zeta_1}\right)\,.\label{hubble_bulk}
\end{equation}

\subsection{Constraints on Alternative Cosmological models with SN}
\begin{table}
\begin{tabular}{c|ccc|c}
\hline
Model & Best fit & parameters&  & $\chi^2_{\textrm{min}}/\textrm{d.o.f.}$\\
\hline
\hline
Volometric   & $H_0=79.05^{+3.85}_{-1.77}$ & $\bar \zeta_0=1.73^{+1.35}_{-0.89}$ & $\bar \zeta_1=0.03^{+0.90}_{-1.26}$& $\chi^2_{\textrm{min}}/\textrm{d.o.f.}=0.971$ \\
\hline
Logotropic    & $H_0=70.25^{+0.69}_{-1.05}$& $\Omega_{m0}=0.28^{+0.15}_{-0.13}$&  $B=0.00^{+0.47}_{-0.20}$ & $\chi^2_{\textrm{min}}/\textrm{d.o.f.}=0.973$ \\
\hline
$\Lambda$CDM    & $H_0=70.04^{+0.68}_{-0.64}$& $\Omega_m=0.28 \pm 0.04$& & $\chi^2_{\textrm{min}}/\textrm{d.o.f.}=0.971$ \\
\hline
\hline
\end{tabular}
\caption{Best fit points for each model obtaining by fitting the theoretical distance moduli with the observed values of the Union  2.1 SN Ia data set.}\label{Table1}
\end{table}

Observations of nearby and distant Type Ia Supernovae (SNe Ia) demonstrated that the expansion of the Universe is accelerating at the current epoch
\cite{Riess:1998cb,Perlmutter:1998np}. This surprising property of the universe was discovered in  1998 by the High-Z Supernova Search Team \cite{Riess:1998cb} and by 
the Supernova Cosmology Project \cite{Perlmutter:1998np}. Since then, a continuos effort in measuring the distance moduli of SNe Ia is being doing by the SCP and currently  
they released the "Union2.1" SN Ia compilation that includes 580 data points that can be use to test cosmological models \cite{Suzuki:2011hu}. In our case, we will test the Logotropic cosmological model 
and the bulk viscous model  by fitting the free parameters of each model with the Union 2.1 data set. 
This is done through a simple $\chi^2$ analysis which compare the measured distance moduli $\mu^{Exp}(z)$ for a supernova at a distance $z$ with the theoretical one $\mu^{Th}$. 
The theoretical distance modulus is computed as
\begin{equation}
  \mu^{Th}(z) = 5 \log \left[ \frac{d_L(z)}{\rm Mpc} \right] +25 \,,
\end{equation}
where the luminosity distance $d_L$ in a spatially flat, homogeneous and isotropic  universe is defined as:
\begin{equation}
  d_L(z) = c(1+z) \int_0^z \frac{dz'}{H(z')}\,,
\end{equation}
$c$ the speed of light. Here the information of each model is included in $H(z)$ which for our cases are given by eqs. \ref{hubble_logo} and \ref{hubble_bulk}. 
Then we define the $\chi^2$ function 
\begin{equation}
\chi^2=\sum_i \left(\frac{\mu^{Th}(z_i)-\mu^{Exp}(z_i)}{\delta \mu^{Exp}_i}\right)^2\,.\label{chi2}
\end{equation}
Here $i$ runs over the 580 points of the Union 2.1 data set and we minimize the $\chi^2$ function over the free parameters of each model. 
Isocurves at 68\% C.L and 90\% C.L for each model are shown in Figs. \ref{cosmology_fit1} and \ref{cosmology_fit2} for the free parameters of each model. For the Logotropic Cosmological model 
this parameters are: the current dark matter energy density $\Omega_m$, the logotropic temperature $B$ and the current Hubble parameter $H_0$. The best points that 
fit the measured distance moduli for this model are shown in Table \ref{Table1}. In the case of the bulk viscous matter-dominated cosmological model, the free parameters are $H_0$ and the
effective dimensionless viscosity coefficient $\bar \zeta_0$ and $\bar \zeta_1$ defined by eq. \ref{zetas}. The best fit parameters for this model are shown in Table \ref{Table1}.
We have performed the same $\chi^2$ analysis for a $\Lambda$-CDM model with the same data points and assuming a flat universe. In this case $\Omega_\Lambda+\Omega_m=1$ and the the number of free parameters is only two, i.e. $\Omega_m$ and $H_0$. The best fit points for $\Omega_m$ and $H_0$ are shown in Table \ref{Table1} as well. 
Note that the central value of $H_0$ for both $\Lambda$CDM and the logotropic cosmological model are pretty similar while for the bulk viscous model is very different. Furthermore, in terms of the goodness of the fit which is measured by $\chi^2_{\textrm{min}}/\textrm{d.o.f.}$ for the three models is very similar and close to one. This implies that the three model, at least in this simple analysis done without taking into account all available data, all of them are able to fit the Union 2.1 data set on SN 1a redshift. This can be see better in Fig. \ref{goodness} where we have plotted $\mu^{Exp}(z_i)$ and the best fit for the two alternative cosmological models we are using in this work. 
Although in both models have a similar fit to the Union 2.1 data, the best value for $H_0$ differs (see Table \ref{Table1}) and the behaviour of 
$\mu^{Th}(z)$ too, as it can be seen in Fig. \ref{goodness}. This differences will be derive a different number of events for the DSNB as it will be seen in the next section.

\begin{center}
%\begin{figure}
\FIGURE{
\includegraphics[angle=0,width=0.9\textwidth]{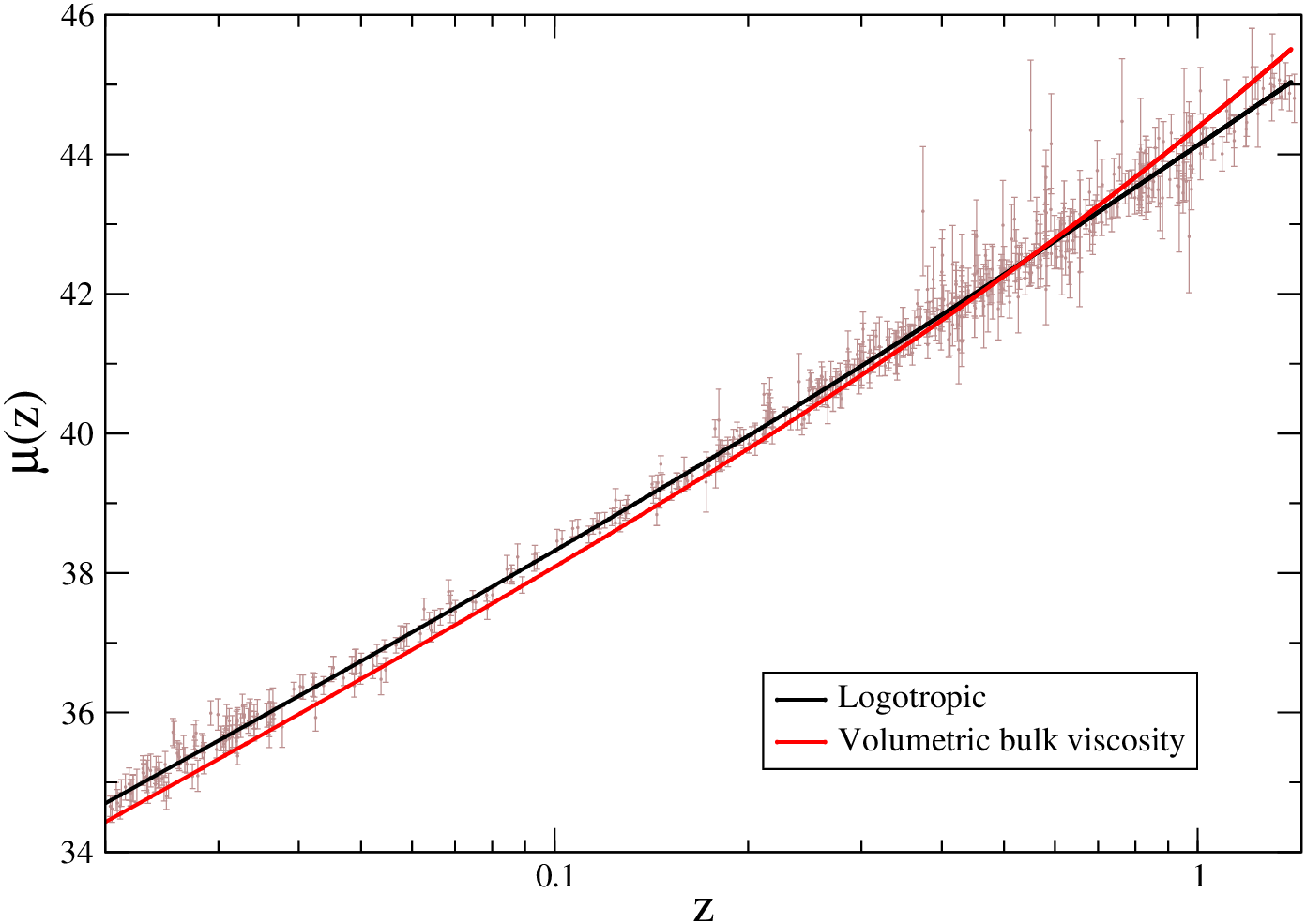}
\caption{Experimental data points from the Union 2.1 data set for the distance moduli and the theoretical distance moduli $\mu^{Th}(z)$ evaluated in the best fit points for each model considered in this work: the logotropic and the bulk viscous cosmological models.}\label{goodness}
}
%\end{figure}  
\end{center}

Before going to the prediction of the number of events of the DSNB flux for this two alternative cosmological models, let us comment some properties of those models that can be inferred from the values obtained through the fit with the Union 2.1 data set. 
\begin{itemize}
\item {\bf Bulk viscous matter dominated  model:} We can estimate the age of the universe for this cosmology. The age of the universe can be computed as 
\begin{equation}
T_{\mbox{Age of Universe}}  = \left| \frac{2}{H_0 \bar{\zeta}_0} \ln \left(1 - \frac{\bar{\zeta}_0}{3 - \bar{\zeta}_1} \right) \right|\,.
\end{equation}
In our case, using the values reported in Table \ref{Table1} we get that the age of the universe will be $T_{\mbox{Age of Universe}}=12.49$ Gyrs in agreement with the age of the oldest globular clusters. 
Furthemore, the best fit predicts an eternally expanding universe that started with a Big Bang followed by a decelerated expansion that later has a smooth transition to an accelerated expansion that occurs at 
\begin{equation}
a_t=\left(\frac{2\bar \zeta_0}{(\bar \zeta_1-1)(\bar \zeta_0+\bar \zeta_1-3)}\right)^{2/(\bar \zeta_1-3)}=0.491\,,
\end{equation}
that correspond to $z_t=1.03$. The best fit parameters ful-fill $\bar \zeta=\bar \zeta_0+\bar \zeta_1 H(z) >0$ for all values of $z$ and thus the second law of thermodynamics is satisfied during all history of the universe. Thus, we may conclude that this model could be a valid alternative to the $\Lambda$-CDM model. Remember that for this model, there is no cosmological constant and all dark sector consist of a perfect fluid and the driving force for the current acceleration of the universe comes from the bulk viscosity of the fluid. 
\item {\bf Logotropic cosmological model:} In this case, we have found that the best fit gives $B=0$ (see Table \ref{Table1}. In this case, the logotropic cosmological model have the same properties as the $\Lambda$-CDM model. As expected, the DSNB flux will be indistinguishable for both models as we will show next.   
\end{itemize}

\section{DSNB and alternative cosmological models: Predictions and constraints}\label{sec3}
Once we have found the best fit points for each model, we can compute the DSNB flux and the expected number of events for both alternative cosmological models that we are considered in this work. 
It is only needed to compute the integral eq. \ref{flujo} but for  $|\frac{dz}{dt}|=(1+z)H(z)$, instead of using $H(z)$ of a $\Lambda$-CDM model (eq. \ref{HLCDM}), we will use 
$H(z)$ in eq. \ref{hubble_logo} for the logotropic universe and $H(z)$ in eq. \ref{hubble_bulk} for the bulk viscous universe. In both cases we use the best fit parameters reported in Table \ref{Table1}.
\begin{center}
\FIGURE{
%\begin{figure}
\includegraphics[angle=0,width=0.9\textwidth]{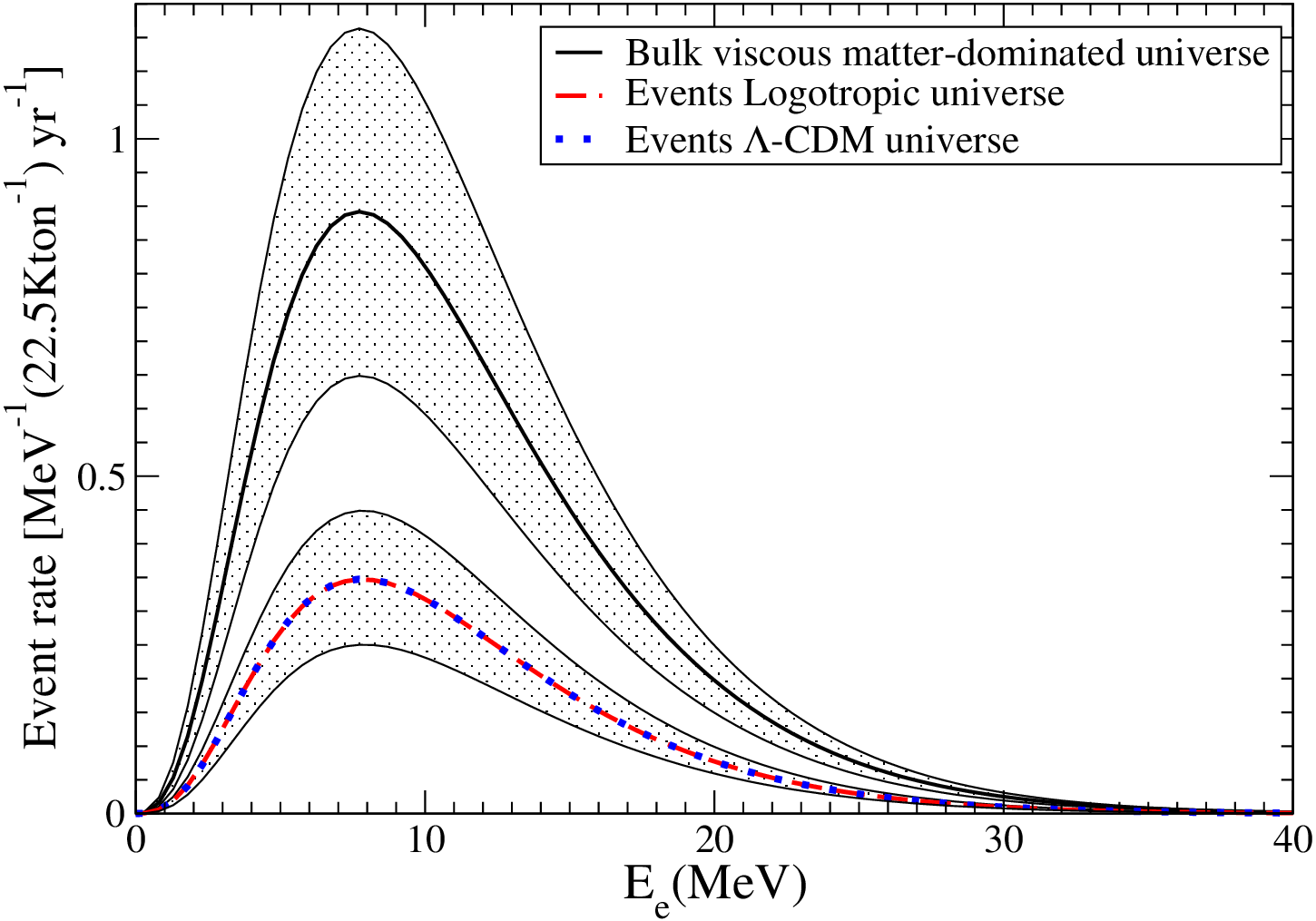}
\caption{The events for a 22 kiloTon detector for $\Lambda$ CDM (dotted line), the logotropic (dashed red line) and the bulk viscous matter dominated  cosmological models for 
the best fit point obtained through a $\chi^2$ analysis of the Union 2.1 data set in SN 1a distance moduli. Grey area correspond to the uncertainty derived from the 
the core collapse SN rate.}\label{model_events}
%\end{figure}  
}
\end{center}
Once $\frac{d\phi^{DSNB}}{dE}$ is computed for both models, we can estimate the number of events for a detector like Super-Kamiokande through eq. \ref{events}. The result are shown in Fig. \ref{model_events} for $\Lambda$ CDM (dotted line), the logotropic (dashed red line) and the bulk viscous matter dominated  cosmological models for 
the best fit point obtained through a $\chi^2$ analysis of the Union 2.1 data set in SN 1a distance moduli reported in Table \ref{Table1}. The grey area in the expected events is obtained by using the upper and lower envelopes that takes into account the scatter in the data of the
core collapse supernova rate $R_{CCSN}$ \cite{Horiuchi:2008jz}.
 
As it can be seen in Fig. \ref{model_events}, the Logotropic and the $ \Lambda$-CDM model prediction are the same, while the number of events for the bulk viscous matter dominated  model are very different, actually, it predicts $\sim 3$ times more events. Thus, the inevitable future detection of the DSNB performed by megaton detectors 
will help to rule out alternative models to the $\Lambda$-CDM model that fit SN 1a data or other cosmological data. 
The high number of events is because the predicted DSNB flux for a bulk viscous matter dominated  universe is higher than the $\Lambda$-CDM model. 

\begin{center}
%\begin{figure}
\FIGURE{
\includegraphics[angle=0,width=0.9\textwidth]{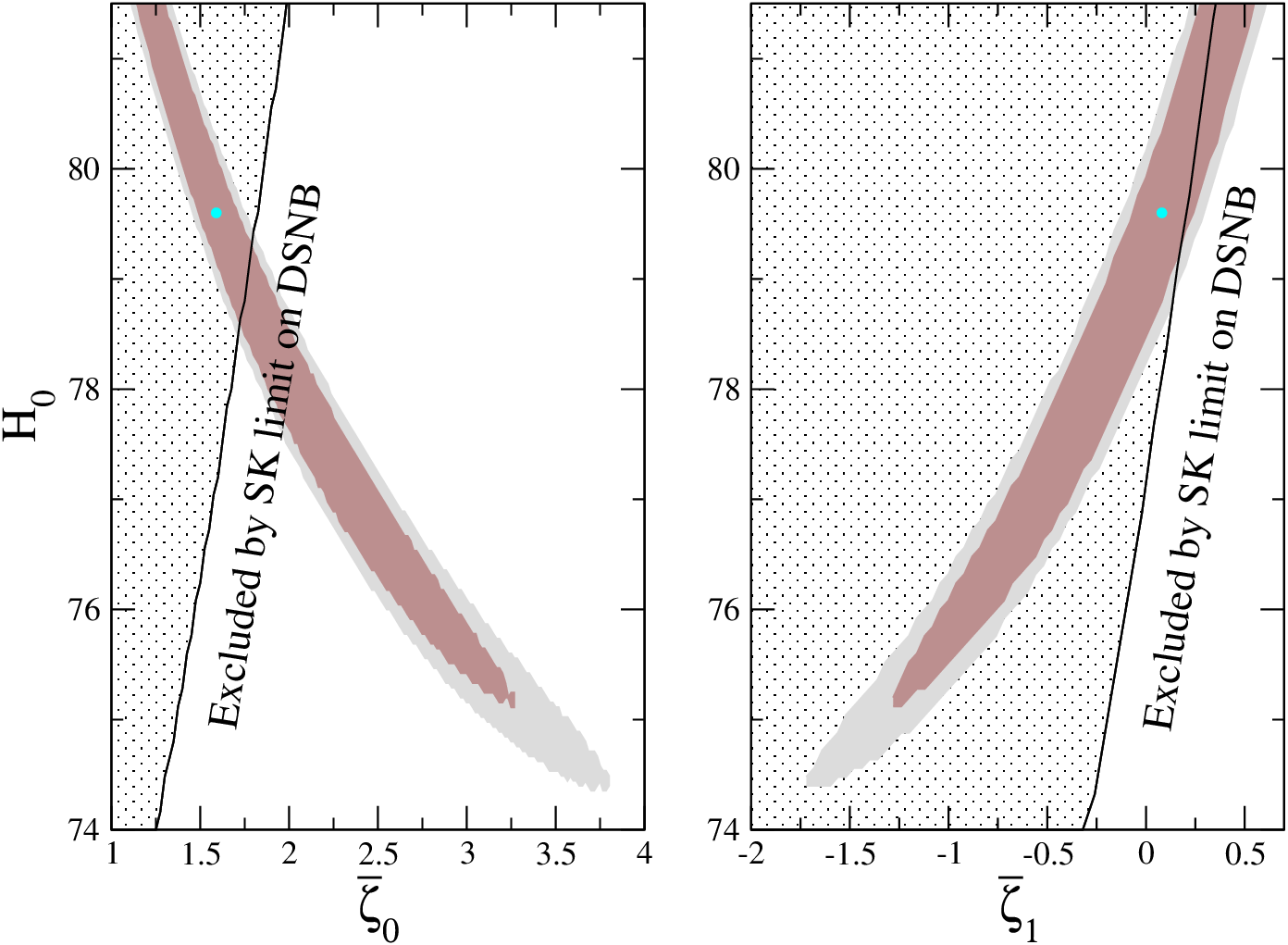}
\caption{Allowed regions for the $H_0,\bar \zeta_0$ and $\bar \zeta_1$ constrained with the Union 2.1 data set at $68\%$ and $90\%$ C.L. and the excluded region obtained by imposing that $\phi^{DSNB}(H_0,\Omega_m) < 1.2~ \bar \nu_e \rm{cm}^{-2} \rm{sec}^{-1}$ for $E_\nu > 19.3$ MeV.}\label{limit_viscoso}
%\end{figure}  
}
\end{center}

By using the current limit set by Super-Kamiokande, we can find the values of $H_0,\bar \zeta_0$ and $\bar \zeta_1$ that predicts $\phi^{DSNB}(H_0,\bar \zeta_0,\bar \zeta_1) < 1.2~ \bar \nu_e \rm{cm}^{-2} \rm{sec}^{-1}$ for $E_\nu > 19.3$ MeV. The allowed region for the parameters of the bulk viscous model that satisfies this constraint are shown in Fig. \ref{limit_viscoso}. In the same figure we have included the allowed regions for the $H_0,\bar \zeta_0$ and $\bar \zeta_1$ constrained with the Union 2.1 data set at $68\%$ and $90\%$ C.L.
We can see that the Super-Kamiokande limit excludes a region that is allowed by the distance moduli for this specific cosmological model. 
Thus, we have shown that for a particular cosmological model, the Super-Kamiokande limit on DSNB can constrain the set of parameters complementary to the limits that can be set for instance for the redshift of supernovas.

\section{Conclusions}\label{sec4}
In this work we have computed the diffuse supernova neutrino background by changing the Hubble parameter $H(z)$ for three different models: a $\Lambda$-CDM model, a model with a fluid with a logotropic equation of state and a model with no cosmological constant but with a fluid that has a bulk viscosity that explain the current acceleration of the universe. 
The free parameters of each model where fixed by fitting the distance moduli with the ``Union 2.1'' data set. Both the logotropic and the bulk viscous matter dominated  models fits with the same degree of accuracy the data and all them have similar $\Delta \chi^2/\rm{d.o.f}$ (See Table \ref{Table1}). 
Then we estimate the expected number of events and spectra for a detector like Super-Kamiokande by using the best fit parameters of each model and we have found that in comparison with the $\Lambda$-CDM model, the logotropic model predicts the same number of events, but the volumetric bulk viscus fluid 
have a different prediction: the number of events is considerable bigger. The reason of this discrepancy can be seen in Figs. \ref{goodness} and in Table \ref{Table1}:
the fit gives a bigger value of $H_0$ and the behaviour of $\mu(z)$ differs from the case of a $\Lambda$-CDM model. The predicted DSNB is considerable bigger 
and thus, given the current limit on the DSNB flux we can constraint the free parameters of this viscous model by demanding that the predicted flux be smaller that the Super-Kamionade limit, i.e. $\phi^{DSNB}(H_0,\bar \zeta_0,\bar \zeta_1) < 1.2~ \bar \nu_e \rm{cm}^{-2} \rm{sec}^{-1}$ for $E_\nu > 19.3$ MeV. It is interesting that the allowed values on the free parameters of this cosmological model, namely  $H_0, \bar \zeta_0,\bar \zeta_1$, obtained by fitting the Union 2.1 data can be 
constrained by this limit on the non observation of the DSNB flux done by Super-Kamiokande. This is illustrated in Fig. \ref{limit_viscoso}.

Furthermore, the Super-Kamiokande limit implies, within a  $\Lambda$ Cold Dark Matter model, that the universe should be expanding with $H_0 > 21.5 ~\rm{Km/sec/Mpc}$ independently of the content of dark matter $\Omega_m$. This can be seen in Fig. \ref{LimitCDM}.

Then we may conclude that the present limit set by Super-Kamiokande on the detection of the diffuse supernova neutrino background is an alternative way of constraining cosmological models such as a bulk viscous matter-dominated universe.  Wr conclude that future detection of DSNB will be of great help in order to test the expansion of the universe for small redshifts.

\section{Acknowledgments}
This work was partially support by CONACYT project CB-259228, Conacyt-SNI and DAIP project.

\end{document}